\documentclass[journal]{IEEEtran}
\usepackage[T1]{fontenc}
\usepackage{float}
\usepackage{amsmath}
\usepackage{amsfonts}
\usepackage{mathtools}
\usepackage{amsfonts}
\IEEEoverridecommandlockouts
\usepackage{amssymb}
\usepackage{graphicx}
\usepackage{epstopdf}
\usepackage{makeidx}
\usepackage{csquotes}
\usepackage{graphicx}
\usepackage{cite}
\usepackage{xcolor}
\usepackage{subcaption}
\usepackage{graphicx}
\usepackage{balance}
\begin{document}
%
\title{Beyond Diagonal RIS for 6G Non-Terrestrial Networks: Potentials and Challenges}

\author{Wali Ullah Khan,~\IEEEmembership{Member,~IEEE,} Asad Mahmood,~\IEEEmembership{Student Member,~IEEE,} Muhammad Ali Jamshed,~\IEEEmembership{Senior Member,~IEEE,} Eva Lagunas,~\IEEEmembership{Senior Member,~IEEE,} Manzoor Ahmed, Symeon Chatzinotas,~\IEEEmembership{Fellow,~IEEE,}
\thanks{Wali Ullah Khan, Asad Mahmood, Eva Lagunas, and Symeon Chatzinotas are with the Interdisciplinary Center for Security, Reliability and Trust (SnT), University of Luxembourg, 1855 Luxembourg City, Luxembourg (e-mails: \{waliullah.khan, asad.mahmood, eva.lagunas, symeon.chatzinotas\}@uni.lu).
Muhammad Ali Jamshed is with the College of Science and Engineering, University	of Glasgow, UK (e-mail: muhammadali.jamshed@glasgow.ac.uk).
Manzoor Ahmed is with the School of Computer and Information Science and also with the Institute for AI Industrial Technology Research, Hubei Engineering University, Xiaogan City, 432000, China  (e-mail: manzoor.achakzai@gmail.com).

}}%

\markboth{Submitted to IEEE
}
{Shell \MakeLowercase{\textit{et al.}}: Bare Demo of IEEEtran.cls for IEEE Journals} 

\maketitle

\begin{abstract}
Reconfigurable intelligent surface (RIS) has emerged as a promising technology in both terrestrial and non-terrestrial networks (NTNs) due to its ability to manipulate wireless environments for better connectivity. Significant studies have been focused on conventional RIS with diagonal phase response matrices. This simple RIS architecture, though less expensive, has limited flexibility in engineering the wireless channels. As the latest member of RIS technology, beyond diagonal RIS (BD-RIS) has recently been proposed in terrestrial setups. Due to the interconnected phase response elements (PREs), BD-RIS significantly enhances the control over the wireless environment. This work proposes the potential and challenges of BD-RIS in NTNs. We begin with the motivation and recent advances in BD-RIS. Subsequently, we discuss the fundamentals of BD-RIS and NTNs. We then outline the application of BD-RIS in NTNs, followed by a case study on BD-RIS enabled non-orthogonal multiple access low earth orbit satellite communication. Finally, we highlight challenges and research directions with concluding remarks.
\end{abstract}


\IEEEpeerreviewmaketitle

\section{Introduction} 
Fifth-generation wireless networks have primarily focused on optimizing transceiver designs to manage the uncontrollable wireless environment. However, as we move into the realm of sixth-generation (6G) and beyond networks, there is a shift towards new emerging concepts such as non-terrestrial networks (NTNs) and the coexistence of terrestrial and non-terrestrial networks. While this integration offers numerous benefits, it also presents the challenge of interference when these networks share the same spectrum resources \cite{9861699}. To address this issue, one possible concept might be manipulating the wireless environment itself, thanks to the innovative technology of reconfigurable intelligent surfaces (RIS). RIS consists of reconfigurable elements capable of altering the propagation environment to enhance the channel gain of the received signals, thus enhancing the spectral efficiency of wireless communications. While the advantages of conventional diagonal RIS (CD-RIS) have been validated in various applications, it relies on a simple diagonal phase response matrix (PRM), where each element independently controls the phase of the incident signal without any cooperation with other elements \cite{10396846}. This approach has some shortcomings: it only allows phase control, limiting passive beamforming (BF) capabilities, and it confines signal reflection to a single direction, thus limiting wireless coverage.

\begin{figure*}[!t]
\centering
\includegraphics[width=1\textwidth]{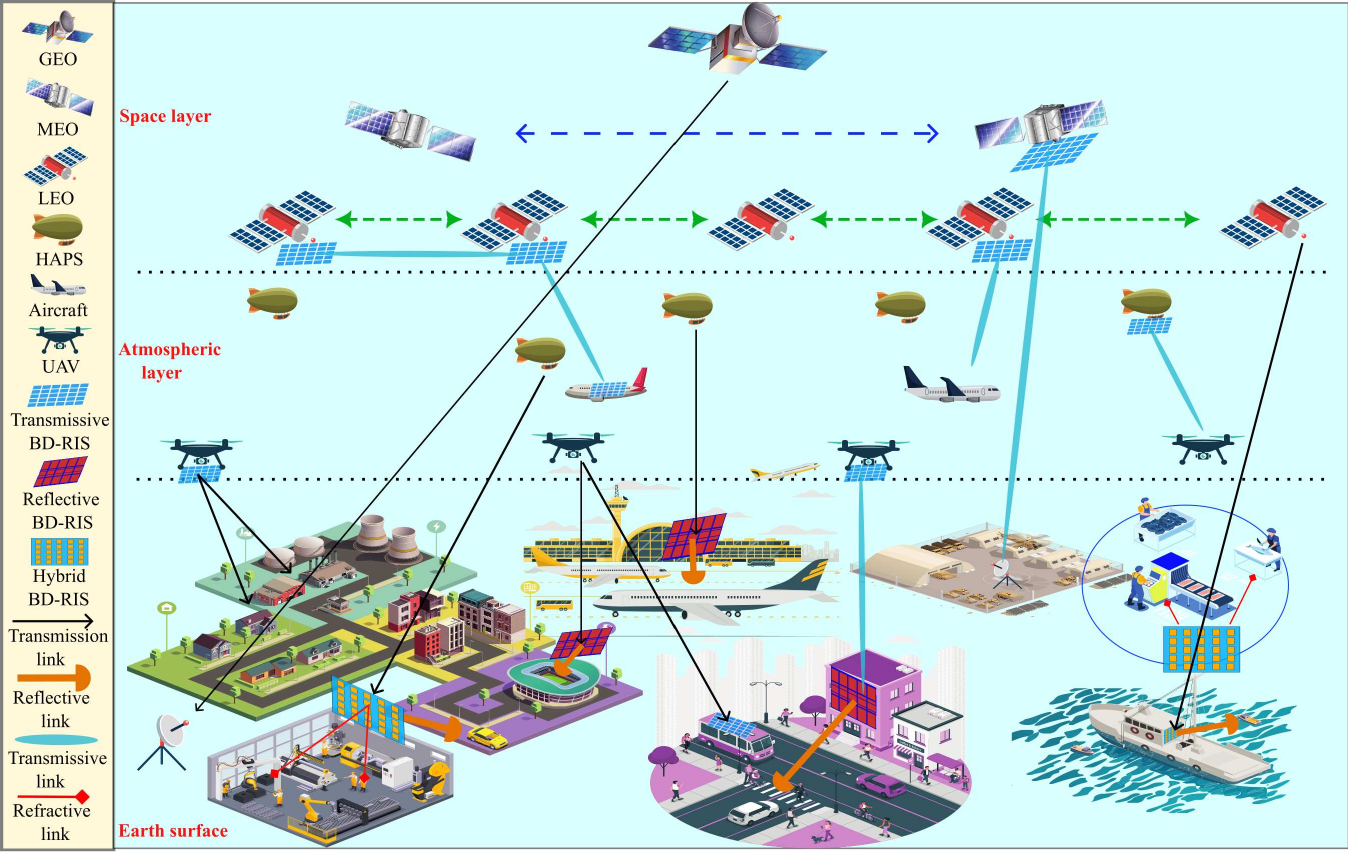}
\caption{BD-RIS enabled 6G NTNs which consist of three layers, i.e., space layer, atmospheric layer, and earth surface. }
\label{blocky}
\end{figure*}

To address these limitations, a new member of RIS technology called beyond diagonal RIS (BD-RIS) has recently merged, which introduces connections among phase response elements (PREs) that significantly enhance control over the reflected signals \cite{10197228}. Moreover, the mathematical model for BD-RIS moves beyond the constraints of diagonal matrices, allowing for more sophisticated wave manipulations. Depending on how the PREs are connected, various BD-RIS architectures can be classified as single-connected, fully-connected, and group-connected BD-RIS, respectively. This leads to different characteristics of PRMs, i.e., block diagonal matrix for group-connected, unitary matrix for fully-connected, and diagonal for single-connected architecture. Note that single-connected is a special case of group-connected such that each group consists of a single PRE. Fig. \ref{blocky} envision a broad range of BD-RIS uses in 6G NTNs. As can be seen in the figure, BD-RIS can be used as transmissive, reflecting, and hybrid modes. While it is true that reflective and hybrid BD-RIS are typically deployed to enhance communication when the line-of-sight (LoS) link is blocked, transmissive BD-RIS can be equipped with non-terrestrial platforms to actively control and optimize the signal propagation, even when the LoS is available\footnote{The transmissive BD-RIS offers more advantages over the reflective BD-RIS for two main reasons. First, the reflective BD-RIS has self-interference because the feed antenna (transmitter) and receiver are on the same side, causing interference between the transmitted and reflected signals. On the other hand, the transmissive BD-RIS has the feed antenna and receivers on the opposite side, so there is no self-interference. Second, unlike reflective BD-RIS, we can design transmissive BD-RIS with enhanced operational bandwidth and aperture efficiency.}. By doing this, the signals can be dynamically shaped and steered not only to improve coverage in areas where the LoS is obstructed but also to enhance overall signal strength and quality in various environments. This approach allows for more flexible and efficient use of non-terrestrial resources, particularly in scenarios where ground-based BD-RIS deployment alone might not address all the communication challenges. Moreover, one of the main benefits of transmissive BD-RIS is performs as a transmitter without necessitating complex signal processing, setting it apart from conventional large-antenna array systems that rely on complicated and energy-hunger RF modules, thereby incurring high hardware costs.
\subsection{Related Works} 
Recently, some works have investigated the performance of BD-RIS in 6G wireless networks. The research work of \cite{10197228} have optimized phase response of BD-RIS in multiple input multiple output system to enhance the received signal gain. In \cite{li2024channel}, Li {\em et al.} have proposed channel estimation and BF design for multi-antenna BD-RIS aided wireless systems to investigate mean square error, spectral efficiency and sum rate. In \cite{10319662}, the authors have derived a closed-form solution for received channel gain maximization of BD-RIS aided multi-user system by optimizing the active and passive BF. In \cite{esmaeilbeig2024beyond}, Esmaeilbeig {\em et al.} have adopted alternating optimization to maximize the signal-to-noise ratio in BD-RIS aided integrated sensing and communication systems. In \cite{10158988}, Li {\em et al.} have computed the scaling law for received signal and maximized the sum rate of multi-user networks by designing an efficient algorithm for multi-sector BD-RIS. In \cite{mahmood2023joint}, Mahmood {\em et al.} have optimized communication and computational resources to reduce the system latency in BD-RIS equipped unmanned aerial vehicle based mobile edge computing. In \cite{10155675}, Nerini {\em et al.} have derived a closed-form solution for designing efficient precoding at the transmitter, combining at the receiver, and BF at BD-RIS to maximize the average received sum power of the system. In \cite{10288244}, Soleymani {\em et al.} have optimized transmit BF and phase response (PR) to improve the performance of BD-RIS aided multi-cell rate splitting multiple access networks. In \cite{10495009}, Liu {\em et al.} have jointly designed linear filter, transmit BF, and PR to maximize the sum rate of BD-RIS aided integrated sensing and communication. Moreover, Zhou {\em et al.} \cite{10364738} have minimized the power consumption and maximized energy efficiency in BD-RIS aided wireless networks by optimizing the transmit BF and BD-RIS PR. In \cite{9913356}, the authors have jointly optimized the transmit precoder and PR to maximize the sum rate of BD-RIS aided multi-user systems. The detailed comparison of the existing works is summarized in Table \ref{Comp}.

\begin{table*}\centering
\begin{tabular}{ |p{1.1cm}|p{1cm}|p{0.8cm}|p{3.5cm}|p{4cm}|p{5cm}| }
 \hline
 Reference& Year & NTNs & Objective & Findings  & Proposed solution\\
 \hline\hline
 \cite{10197228}   &  2023    &  $\bigtimes$ & Discrete PR design for BD-RIS &  Received signal power &  Used alternating optimization approach for discrete PR design.\\\hline
 \cite{li2024channel}   & 2024    & $\bigtimes$ & CE and BF design & Mean square error, SE, SR & Used least square method for CE and efficient algorithm for BF design.\\\hline
 \cite{10319662} &   2024  & $\bigtimes$   &Active and passive BF design & SR, effective channel gain and time complexity & Exploited two-stage low-complex joint active and passive BF at BS and BD-RIS.\\\hline
 \cite{esmaeilbeig2024beyond}  &2024& $\bigtimes$ &  PR design for BD-RIS & Signal-to-noise ratio & Adopted an alternating optimization to jointly optimize the $\Phi$ for integrated sensing and communication. \\\hline
 \cite{10158988}  & 2023 & $\bigtimes$ &  BF design for multi-sector BD-RIS & SR & Derived the scaling law for received signal and Designed an efficient algorithm for multi-sector BD-RIS.\\\hline
\cite{mahmood2023joint} &   2024  & $\bigtimes$ & BD-RIS deployment, computational and communication resources & Latency, max-min rate, task segmentation and resource allocation & Proposed a joint iterative optimization approach for BD-RIS deployment, computational and communication resources.\\\hline
 \cite{10155675} & 2024  & $\bigtimes$   & Precoding, combining, and BF design & Average received sum power and computational complexity & Derived a closed-form solution for precoding at transmitter, combining at receiver and BF at BD-RIS. \\\hline
 \cite{10288244} & 2024  & $\bigtimes$ & BF and PR design & Fairness rate and fairness EE & Proposed majorization minimization
and alternating optimization to iteratively solve transmit BF and PR problem for fairness rate and EE improvement.\\
 \hline
 \cite{10495009} & 2024  & $\bigtimes$ & Linear filter, transmit BF and BD-RIS PR design & SR & Used majorization-minimization and penalty-dual decomposition methods to solve the problem of joint linear filter, transmit BF, and PR design.\\
 \hline
 \cite{10364738} & 2023  & $\bigtimes$ & Transmit BF and BD-RIS PR design & SR, EE, power consumption & Utilized a uni-fold optimization approach to solve the problem of total power consumption minimization and EE maximization.\\
 \hline
 \cite{9913356} & 2023  & $\bigtimes$ & Transmit precoding and BD-RIS PR design & SR & Employed fractional programming theory to solve the transmit precoding and PR problem for sum rate maximization.\\
 \hline
 \multicolumn{6}{l}{\textit{Ref-Reference, CE-Channel estimation, BF-Beamforming, SE-Spectral efficiency, EE-Energy efficiency, SR-Sum rate.}}
\end{tabular}\caption{Summary of different works on BD-RIS aided 6G wireless networks.}
\label{Comp}
\end{table*}

Although some research has been done on BD-RIS aided terrestrial networks, there is no existing literature that considers BD-RIS in NTNs. Motivated by this, we study the potential and challenges of BD-RIS in 6G NTNs. The article first discusses the fundamentals of NTNs and BD-RIS technology. Then, it explains the applications of BD-RIS in NTNs. Subsequently, it provides a case study on BD-RIS aided LEO satellite communication using downlink non-orthogonal multiple access (NOMA). Finally, this work highlights challenges and future research directions in BD-RIS aided 6G NTNs.

\section{Non-Terrestrial Networks}
NTNs is a sophisticated and diverse network that consists of satellite platforms and aerial platforms and ground terminals. It combines the wide coverage of satellite networks in the space layer and the adaptability of aerial networks in the atmospheric layer, as illustrated in Fig. \ref{blocky}. 
\subsection{Atmospheric Layer}
The aerial network in the atmospheric layer primarily consists of unmanned aerial vehicles (UAVs) and high altitude platform stations (HAPSs) \cite{9861699}. The UAVs function at varying heights, spanning from hundreds to thousands of meters, and have the capability to cover distances of many kilometers (km). The operation is straightforward, the deployment is convenient, and the mobility is versatile. It has the capability to establish network connections and transmit signals through relays. However, the device's operational duration varies from a few minutes to a few hours due to its limited battery capacity. On the other hand, HAPSs are situated in the stratosphere at an altitude ranging from 20 km to 50 km above the Earth's surface. These platforms tend to remain in a fixed position. A network is established among many HAPSs by mmWave or THz connections. Because HAPSs are positioned at high altitudes, each HAPS has an extensive coverage area that can span up to 100 km in circumference when the elevation angle is 10 degrees. Consequently, by reducing the number of HAPSs, it is possible to obtain a broader coverage area and expedite deployment. HAPS communication systems possess the attributes of cost-effectiveness, minimal latency, rapid deployment, and high capacity in contrast to satellite systems.

\begin{figure}[!t]
\centering
\includegraphics[width=0.48\textwidth]{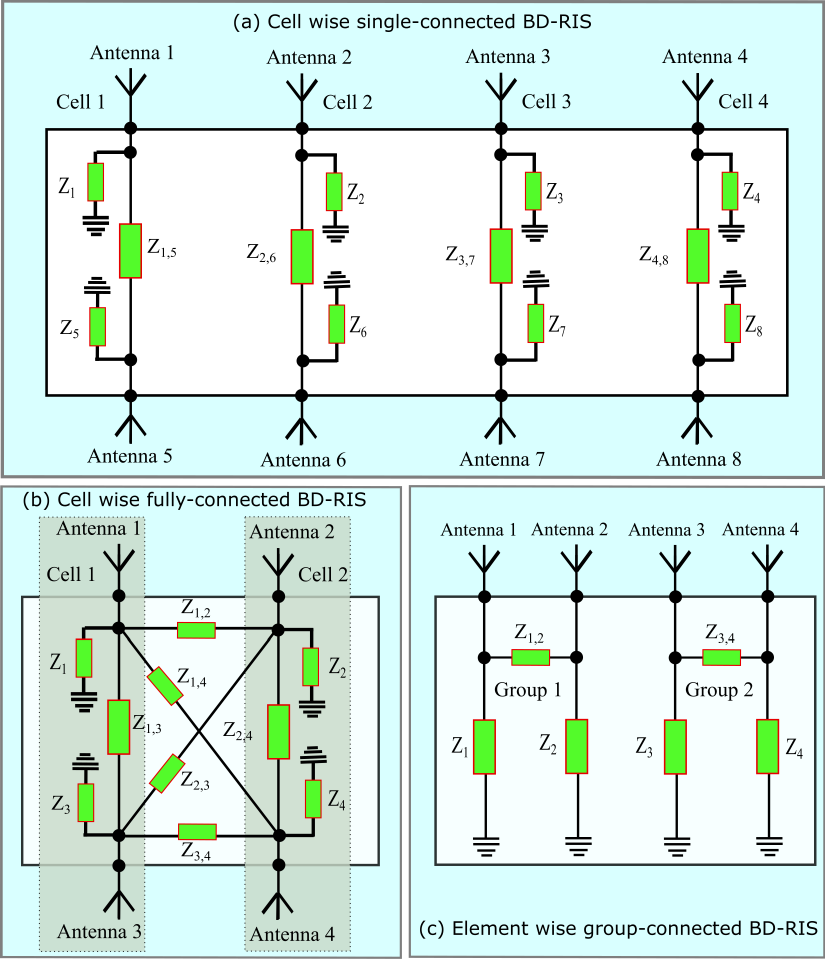}
\caption{Different architecture of BD-RIS: (a) Cell-wise single-connected BD-RIS architecture with 2 cells; (b) Cell-wise fully-connected BD-RIS architecture with 4 cells and 2 groups; (c) Element-wise group-connected BD-RIS with 2 groups.}
\label{architecture}
\end{figure}
The aerial network possesses exceptional flexibility and scalability, making it suitable for a wide range of applications such as emergency communication, network traffic offloading, and temporary communication. For instance, in the event of natural disasters causing the destruction of the terrestrial network, the aerial network can be swiftly deployed to restore the connection. Furthermore, an aerial network exhibits a reduced expenditure and decreased delay compared to a satellite network. Nevertheless, the limitations on energy consumption and weight have resulted in a significant drawback for the aerial network, namely its limited endurance period. Moreover, due to the HAPSs restricted height, it can easily cause significant interference to the terrestrial network.
\subsection{Space Layer}
This layer includes different satellite networks. Based on height, the satellite network has three tiers: geostationary orbit (GEO), medium Earth orbit (MEO), and LEO \cite{9861699}. GEO satellites form the core space-based network, whereas MEO and LEO satellites form the access network that connects to the aerial or terrestrial network. GEO satellites are set at 35,786 km and the transmission requires 0.27 seconds to reach from GEO to the ground terminal. GEO satellites cover up to one-third of Earth's surface. Except for the South and North Poles, three GEO satellites evenly spaced offer global communication. The majority of MEO satellites operate between 8,000 and 20,000 km altitude and they require 0.1 second for signal transmission to the ground terminal. These satellites mostly deliver high-speed data and messages. Many satellites operate exclusively in this orbit for telecommunications and military use. LEO satellites orbit between 500 and 2000 km above sea level. From this orbit, the satellite's signal reaches the ground station in 0.05 seconds. This track is best for low-latency applications.

Thanks to their exceptionally high altitude, a limited number of satellites can provide remarkably wide coverage, unaffected by geographical features. This is particularly crucial for rural regions and maritime communication. Simultaneously, satellites harness solar panels to generate energy, with an average lifespan of approximately 10 years, making them a superb addition to terrestrial communication. These advantages facilitate the growth of satellite networks. As an illustration, in 2023, the United States successfully deployed Viasat-3 satellites with a remarkable capacity of 1 terabits per second (Tbps). Additionally, SpaceX has launched a multitude of LEO Starlink satellites. Nevertheless, the existing satellite network continues to suffer from limitations in transmission capacity and significant latency.

\subsection{Channel Modeling in NTNs and Integration with BD-RIS}
The channel modeling and characteristics in NTNs differ from those in terrestrial networks. For example, compared to terrestrial network channels, the channels in NTNs have challenges of high mobility, high path loss due to the large distance between satellites and ground terminals, Doppler shift caused by the satellite spinning, and signal blockage due to geographical obstructions and atmospheric absorptions. Besides that, the link budget in NTN systems is often constrained, especially for ground terminals with small antennas operating at higher frequencies, where signal attenuation is more prominent. 
BD-RIS can add more degrees of freedom in controlling and optimizing signal reflections, hence its integration with NTNs fits well with these channel properties. In particular, BD-RIS can dynamically change its phase shifts and reflection characteristics to match the fast-changing NTN channels, therefore adjusting to Doppler effects, signal fading, and multipath propagation. Furthermore, by concentrating signal reflections more accurately onto the user's antenna, BD-RIS can improve the link budget by offsetting the great path loss and guaranteeing improved signal strength at the receiver. BD-RIS is a necessary instrument for maximizing NTN channel performance, especially in demanding link budget situations, since it allows one to enhance the effective signal gain.
\begin{figure*}[!t]
\centering
\includegraphics[width=1\textwidth]{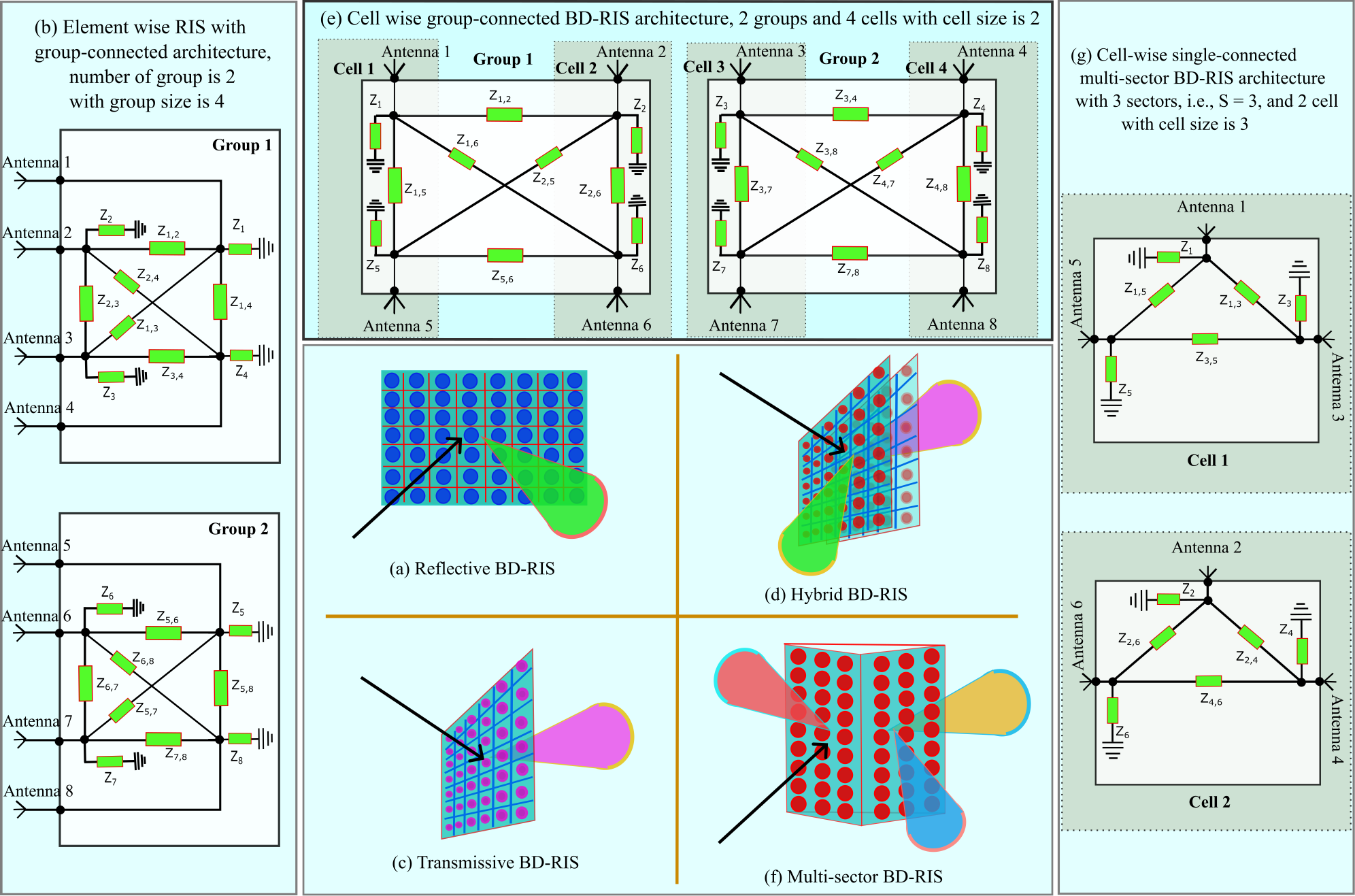}
\caption{Different modes BD-RIS supported by the same circuitry of reconfigurable impedance network: (a) Reflective BD-RIS, (b) Element wise reflective BD-RIS with group-connected architecture with 2 groups, where the group size is 4, (c) Transmissive BD-RIS, (d) Hybrid BD-RIS; (e) Cell wise group-connected BD-RIS architecture with 2 groups and 4 cells, where the size of each cell is 2; (f) Multi-sector BD-RIS; and (g) Cell wise single-connected multi-sector BD-RIS architecture with 3 sectors and 2 cells, where the cell size is 3.}
\label{modes}
\end{figure*}

\section{Preliminaries of BD-RIS} 
In this section, we first study the hardware architectures of BD-RIS, followed by different operating modes of this technology, and then we also discuss its advantages compared to the CD-RIS technology.

\subsection{Hardware Architectures of BD-RIS }
The hardware architectures of BD-RIS can be classified into three main categories: 1) single-connected BD-RIS, 2) fully-connected BD-RIS, and 3) group-connected BD-RIS, as illustrated in Fig \ref{architecture}. In the following, we briefly discuss each category principle and the matrices they produce. 
\subsubsection{Single-connected BD-RIS}
The hardware architecture of BD-RIS in this category consists of independent reconfigurable elements. The PREs on the surface of a single-connected BD-RIS are not interconnected and each element independently manipulates the incident electromagnetic wave \cite{9913356}. A cell-wise\footnote{In this work, we use the terms \lq element' and \lq cell' interchangeably based on the mode of the BD-RIS. Please note that we use \lq element" when we refer to BD-RIS operating in reflecting mode and use \lq cell when referring to BD-RIS operating in hybrid/multi-sector modes, where more than 1 element with uni-directional radiating patterns are back-to-back placed to form a cell.} single-connected hybrid BD-RIS architecture with 4 cells, where each cell contains 2 elements, is illustrated in Fig. \ref{architecture}(a). A hybrid BD-RIS with $K$ PREs will provide diagonal PRMs as ${\bf \Phi}_r=\text{diag}(\phi_{r,1},\phi_{r,2}\dots\phi_{r,K})$ and ${\bf \Phi}_t=\text{diag}(\phi_{t,1},\phi_{t,2}\dots\phi_{t,K})$, where ${\bf \Phi}_r$ is the reflecting PRM and ${\bf \Phi}_t$ is the transmissive PRM. Furthermore, the PREs in each cell must satisfy as $|\phi_{r,k}|^2+|\phi_{t,k}|^2=1$, $\forall k\in K$.
\subsubsection{Fully-connected BD-RIS}
Unlike single-connected BD-RIS, all the PREs in this architecture are interconnected through an impedance network. A cell-wise fully-connected hybrid BD-RIS architecture with 2 cells, where each cell consists of 2 PREs, is shown in Fig. \ref{architecture}(b). The PRMs ${\bf \Phi}_r$ and ${\bf \Phi}_t$ in this architecture are non-diagonal (full scattering matrices), resulting in improved system performance. The PRMs of this architecture satisfies as ${\bf \Phi}_r^H{\bf \Phi}_r+{\bf \Phi}_t^H{\bf \Phi}_t={\bf I}_K$, where ${\bf I}_K$ is an identity matrix. 
\subsubsection{Group-connected BD-RIS}
To reduce the complexity of fully-connected DB-RIS architecture when the number of PREs is very large, group-connected architecture is introduced, which divides PREs into multiple groups. Figure \ref{architecture}(b) illustrates an element-wise group-connected reflective BD-RIS with 2 groups, where each group consists of 2 elements with a fully-connected configuration. In this architecture, each group is fully-connected and all groups constitute a block diagonal matrix such as ${\bf \Phi}_r=\text{blkdiag}({\bf \Phi}_{r,1},{\bf \Phi}_{r,2},\dots,{\bf \Phi}_{r,G} )$ with $G$ is the number of groups. If $\Bar{K}=K/G$ denotes the number of elements in each group, then the PRM of each group satisfies ${\bf \Phi}^H_{r,g}{\bf \Phi}_{r,g}={\bf I}_{\bar{K}}$ with ${\bf \Phi}_{r,g}\in\mathcal C^{\Bar{K}\times \Bar{K}}$. 

\subsection{Operating Modes of BD-RIS in NTNs}
Each architecture of BD-RIS can operate in three modes, including reflective BD-RIS, hybrid BD-RIS, and multi-sector BD-RIS.
\subsubsection{Reflective BD-RIS}
In this mode, the feed antenna (transmitter) and receiver are located on the same side of BD-RIS, resulting in coverage of 180-degree space\footnote{Transmissive BD-RIS, while similar to reflective BD-RIS, refracts signals instead of reflecting them, allowing the signals to pass through surfaces, as depicted in Fig. \ref{modes} (c). This mechanism is helpful for signal penetration through barriers like windows, walls, or partitions. Due to space limitations, we do not discuss transmissive BD-RIS here in detail.}. The incident signal and the reflected signal occur on the same side of the surface, and self-interference results that impact the system performance \cite{10197228}. As can be seen in Fig. \ref{modes}(a), all of the PREs of the BD-RIS are oriented in the same direction in order to enable the reflective mode during transmission. When considering the BD-RIS with reflecting mode, the PRM $\boldsymbol{\Phi}$, which is subject to a unitary constraint, stands out as the defining characteristic. Since this BD-RIS mode can only reflect the signal, therefore $\boldsymbol{\Phi}_r$ is the unitary matrix which satisfies $\boldsymbol{\Phi}_r^H\boldsymbol{\Phi}_r={\bf I}_K$, where $\boldsymbol{\Phi}_r^H$ is the conjugate transpose of $\boldsymbol{\Phi}_r$ and ${\bf I}_K$ denotes an identity matrix. This property ensures that the columns (and rows) of the matrix are orthogonal. In the context of BD-RIS for 6G wireless networks, unitary constraints are applied to the design of these surfaces to ensure efficient signal manipulation without loss of power. Fig. \ref{modes}(b) is an example of element wise reflective BD-RIS with group-connected architecture, where the number of PREs in each group is 4. In this example, the number of groups is $G=2$, where each group consists of $\hat{K}=K/G=4$ elements. The PRM can be defined as $\boldsymbol{\Phi}_r=\text{blkdiag}(\boldsymbol{\Phi}_{r,1},\boldsymbol{\Phi}_{r,2})$ where $\boldsymbol{\Phi}_{r,1}$ and $\boldsymbol{\Phi}_{r,2}$ satisfy $\boldsymbol{\Phi}_{r,1}^H\boldsymbol{\Phi}_{r,1}={\bf I}_4$ and $\boldsymbol{\Phi}_{r,2}^H\boldsymbol{\Phi}_{r,2}={\bf I}_4$. The reflective BD-RIS can be placed on walls, building facades, or ceilings to enhance signal coverage by bouncing the signal from the source to the destination.

\subsubsection{Hybrid BD-RIS}
This mode allows signals that are impinging on one side of the BD-RIS to be partially reflected toward the same side and partially transmitted toward the opposing side, which results in a coverage of 360-degree space. The BD-RIS that operates in single-connected hybrid mode is also referred to as the intelligent omni-surface or the simultaneous transmitting and reflecting BD-RIS. In order to support the hybrid mode, two elements with a unidirectional radiation pattern are placed back to back to form a cell. These elements are then connected to a two-port fully connected reconfigurable impedance network, as shown in Fig. \ref{modes}(d). This ensures that each reflecting element and each transmitting element cover 180-degree space, respectively, in order to achieve 360-degree coverage. In terms of mathematics, a $K$-cell BD-RIS operating in hybrid mode is characterized by two matrices, namely reflecting and transmitting, i.e., $\boldsymbol{\Phi}_r$ and $\boldsymbol{\Phi}_t$. These matrices should satisfy $\boldsymbol{\Phi}^H_r\boldsymbol{\Phi}_r+\boldsymbol{\Phi}^H_t\boldsymbol{\Phi}_t={\bf I}_{K}$. Fig. \ref{modes}(e) provides an example of cell-wise group-connected hybrid BD-RIS with 2 groups and 4 cells, where the group size is $\hat{K}=K/G=8/2=4$ and cell size is 2. In this case, $\boldsymbol{\Phi}_r$ and $\boldsymbol{\Phi}_t$ can be derived as $\boldsymbol{\Phi}_r=\text{blkdiag}(\boldsymbol{\Phi}_{r,1},\boldsymbol{\Phi}_{r,2})$ and $\boldsymbol{\Phi}_t=\text{blkdiag}(\boldsymbol{\Phi}_{t,1},\boldsymbol{\Phi}_{t,2})$. These matrices should satisfy as $\boldsymbol{\Phi}_{r,1}^H\boldsymbol{\Phi}_{r,1}+\boldsymbol{\Phi}_{t,1}^H\boldsymbol{\Phi}_{t,1}={\bf I}_4$ and $\boldsymbol{\Phi}_{r,2}^H\boldsymbol{\Phi}_{r,2}+\boldsymbol{\Phi}_{t,2}^H\boldsymbol{\Phi}_{t,2}={\bf I}_4$.

\begin{table*}[!t] 
\renewcommand{\arraystretch}{1.2} 
\caption{Characteristics of different BD-RIS architectures and modes.}
\centering
\begin{tabular}{|c|c|c|c|}
\hline
Characteristics & Single-connected architecture & Fully-connected architecture & Group-connected architecture  \\
\hline\hline
Number of groups & $K$ & 1 & $G$ \\
\hline
Group dimension & 1 & $K$ & $\hat{K}$ \\
\hline
Elements/group & 1 & $K^2$ & $\hat{K}^2$\\
\hline
Number of Non-zero elements & $K$ & $K^2$ & $G\hat{K}^2$ \\
\hline
Reflective mode & $|\phi_{r,k}|^2=1$ & ${\bf \Phi}^H_{r}{\bf \Phi}_{r}={\bf I}_K$ & ${\bf \Phi}^H_{r,g}{\bf \Phi}_{r,g}={\bf I}_{\hat{K}}$  \\
\hline
Transmissive mode & $|\phi_{t,k}|^2=1$  & ${\bf \Phi}^H_{t}{\bf \Phi}_{t}={\bf I}_K$ & ${\bf \Phi}^H_{t,g}{\bf \Phi}_{t,g}={\bf I}_{\hat{K}}$  \\
\hline
Hybrid mode &$|\phi_{r,k}|^2+|\phi_{t,k}|^2=1$ & ${\bf \Phi}^H_{r}{\bf \Phi}_{r}+{\bf \Phi}^H_{t}{\bf \Phi}_{t}={\bf I}_K$  & ${\bf \Phi}^H_{r,g}{\bf \Phi}_{r,g}+{\bf \Phi}^H_{t,g}{\bf \Phi}_{t,g}={\bf I}_{\hat{K}}$ \\
\hline
Hardware complexity  &R$^\star$: $K$, H$^\star$: $(3/2)K$, M$^\star$: $(S+1)K/2$ & R$^\star$,H$^\star$,M$^\star$: $(K+1)\frac{K}{2}$  & R$^\star$,H$^\star$,M$^\star$: $(\frac{K}{G}+1)\frac{K}{2}$ \\
\hline
\multicolumn{4}{l}{$^{\star}$R-Reflective, H-Hybrid, M-Multi sector.}
\end{tabular}   
\label{Table2} 
\end{table*} 
\subsubsection{Multi-Sector BD-RIS}
 This mode is an extension or a more comprehensive version of the hybrid mode. In multi-sector BD-RIS, the entire area is partitioned into multiple sectors, where the number of sectors must be greater than or equal to 2, as shown in \ref{modes}(f). Signals that come into contact with one sector of the BD-RIS can be partially redirected back to the same sector and partially dispersed toward the remaining sectors. Let us assume that $S$ is the number of sectors that constitute multi-sector BD-RIS; then, to facilitate the multi-sector mode of BD-RIS, each cell consists of $S$ PREs positioned at the edges of an $S$-sided polygon. These PREs have a unidirectional radiation pattern such that each PRE can cover the $1/S$ area of the sector, preventing any overlap between adjacent sectors. The $S$ PREs are connected to a $S$-port fully connected reconfigurable impedance network. Therefore, the multi-sector mode of BD-RIS can cover 360-degree space, similar to the hybrid BD-RIS. However, multi-sector BD-RIS offers greater performance improvements compared to the hybrid mode due to the utilization of high-gain PREs with narrower beamwidth, which can cover 180-degree space. The BD-RIS with multi-sector mode can be mathematically described by a set of $S$ matrices, $\boldsymbol{\Phi}_s\in \mathbb C^{K/S\times K/S}$, $s=\{1,2,\dots,S\}$, which fulfilling $\sum_{s=1}^S\boldsymbol{\Phi}_s^H\boldsymbol{\Phi}_s={\bf I}_{K/S}$. To further understand the concept of this mode, Fig. \ref{modes}(g) provides cell-wise single-connected multi-sector BD-RIS with 3 sectors and 2 cells, where the number of BD-RIS elements in each cell is 3. The PRMs should satisfy as $\boldsymbol{\Phi}_1^H\boldsymbol{\Phi}_1+\boldsymbol{\Phi}_2^H\boldsymbol{\Phi}_2+\boldsymbol{\Phi}_3^H\boldsymbol{\Phi}_3={\bf I}_{2}$. 

The detailed characteristics of different BD-RIS hardware architectures and operating modes are summarised in Table \ref{Table2}.

\subsection{Advantages of BD-RIS in NTNs}
BD-RIS provides substantial benefits compared to its conventional counterpart due to the connection between elements and the full scattering matrix. Some of the key advantages of BD-RIS over conventional RIS are discussed in the following:
\subsubsection{Advance BF and High Flexibility}
The functionality of conventional RIS is restricted to controlling the phase of the incident signal. As a result, its capacity to manipulate passive BF is limited, leading to a degradation in performance. BD-RIS provides inter-element connections; however, this comes at the cost of increased circuit complexity; its mathematical description is not restricted to diagonal PRMs. This makes for more precise and adaptive BF, which boosts signal quality and strength in many directions. Moreover, the flexibility of BD-RIS for manipulating diagonal and non-diagonal elements of the PRM significantly improves the system performance of various communication scenarios in 6G NTNs.

\subsubsection{Enhanced Adaptability to Challenging Environments}
Due to its limited BF capability, CD-RIS is only effective in non-terrestrial scenarios with relatively stable and predictable links. In contrast, BD-RIS significantly enhances adaptability to more dynamic and challenging environments in NTNs. BD-RIS dynamically adjusts its configuration to accommodate changes in user positions. This ensures optimal signal reception and transmission even as users move, enhancing overall connectivity and reducing signal degradation. Moreover, in environments with moving obstacles or dynamic blockages, BD-RIS can effectively counteract signal interference caused by these obstructions. Its advanced reconfiguration capabilities allow it to maintain reliable communication links despite mobile barriers. Furthermore, the design of BD-RIS adapts to changing network demands, including varying traffic loads and shifting service requirements. This flexibility ensures consistent performance across different scenarios, accommodating both peak and off-peak traffic conditions. 
\subsubsection{Better Interference Management}
Unlike CD-RIS, which offers limited interference mitigation by adjusting the phase of incident signals, BD-RIS provides significantly enhanced interference management capabilities, particularly in NTNs. More specifically, BD-RIS employs advanced signal processing techniques and sophisticated configurations to manage interference more effectively than its conventional counterpart. In NTNs, where signal paths can be complicated because of satellites and other high-altitude platforms, BD-RIS works very well because it changes both the phase and amplitude of signals as they come in. This capability allows BD-RIS to adapt to varying interference conditions and mitigate interference from multiple sources, including both terrestrial and non-terrestrial transmissions.

\subsubsection{360-Degree Coverage}
BD-RIS provides complete 360-degree coverage, unlike CD-RIS, which typically offers limited coverage of up to 180 degrees. BD-RIS operates in hybrid and multi-sector modes, allowing it to extend its coverage range effectively. The multi-sector mode leverages smaller beam widths and the higher element gain of each BD-RIS element to enhance channel gain. This approach not only broadens the communication range but also ensures full-space coverage, making BD-RIS ideal for environments requiring extensive and uniform signal distribution. The ability to deliver 360-degree coverage represents a significant advancement over CD-RIS configurations, providing more reliable and widespread communication in various NTN settings. 
 

\section{Applications of BD-RIS in NTNs}
Well-established research on CD-RIS clearly demonstrates its use cases and applications in diverse fields, including, but not limited to, wireless power transfer, coverage improvements, and wireless sensing. In the following sections, we showcase, for the first time, some of the potential applications of BD-RIS in an NTN-enabled communication scenario.

\subsection{Seamless and Ultra Precision Indoor Positioning}

It is well-known that wireless signals based on NTN are unable to penetrate indoor environments effectively, as they are weakened and scattered by roofs, walls, and other structures. Moreover, the availability of line-of-sight (LoS) signals greatly impacts positioning accuracy. These limitations necessitate a separate infrastructure for indoor positioning applications, thereby increasing overall costs and energy requirements. Since BD-RIS can flexibly reflect signals in multiple controlled directions, it can play a pivotal role in providing an effective conduit for NTN signals to overcome penetration losses caused by building structures. Furthermore, compared to CD-RIS, BD-RIS offers enhanced BF capabilities, thus improving a network's ability to achieve ultra-precise localization of devices. In conclusion, these additional features of BD-RIS can be crucial in developing a new class of indoor positioning systems that do not require any additional infrastructure, costs, or energy resources.

\subsection{Improved IoT Connectivity}

The Internet of Things (IoT) has been a familiar concept for quite some time, but its popularity surged with the integration of machine-type communication (mMTC) as a feature of 5G. IoT encompasses a diverse range of applications, spanning from smart home automation to monitoring deforestation activities. Particularly in areas with limited or no cellular coverage, connecting these IoT devices to the internet poses a significant challenge. To address such obstacles, the 3GPP release 17 incorporates NTN into traditional Terrestrial Networks (TN), enabling NTNs to serve as a bridge between the internet and these IoT devices. Despite playing a critical role in overcoming these limitations, NTN introduces its own set of challenges, such as coverage (while considering smart home indoor environment applications in remote/rural areas) and power consumption. Similarly, the enhanced capabilities of BD-RIS, which include reflecting signals in multiple controlled directions and improved BF, have the potential to mitigate these barriers and establish a seamless connection between NTN and IoT devices.

\subsection{Extended Coverage for NTN-aided Terrestrial Users}

One of the core applications of CD-RIS, for which it was originally developed, is to provide extended coverage, overcome blind spots, and ensure seamless connectivity for terrestrial users. In the context of NTN-aided communication, CD-RIS has shown great potential in improving the performance of cellular users with weak or no connectivity to the base station. For instance, as demonstrated in \cite{jamshed2024synergizing}, CD-RIS aided UAVs (a specific use case of NTN) have clearly illustrated the usefulness of CD-RIS in NTN systems. Although CD-RIS aided NTN has shown significant potential in enhancing the performance of terrestrial users, the limitations associated with CD-RIS prevent us from fully benefiting from NTN. As discussed in the preceding sections, BD-RIS clearly surpasses CD-RIS by offering greater flexibility in reflecting signals in multiple controlled directions. This capability provides more extended coverage compared to CD-RIS and plays a significant role in increasing the performance of NTN.
\section{Case Study: BD-RIS aided NOMA LEO Satellite Communications}
In this section, we provide a case study on BD-RIS aided NTNs. We will first discuss the considered system model and proposed optimization. Then, we provide the numerical results and their discussion. 
\begin{figure*}[!t]
\centering
\begin{subfigure}{.33\textwidth}
  \centering
  \includegraphics[height=0.2\textheight]{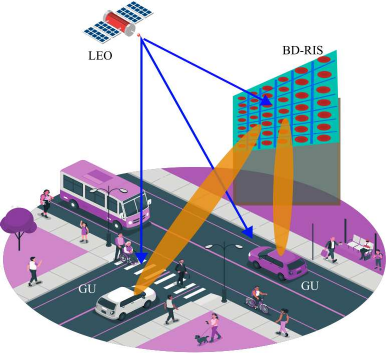}  
  \caption{}
  \label{fig:sub1}
\end{subfigure}%
\begin{subfigure}{.33\textwidth}
  \centering
  \includegraphics[height=0.2\textheight]{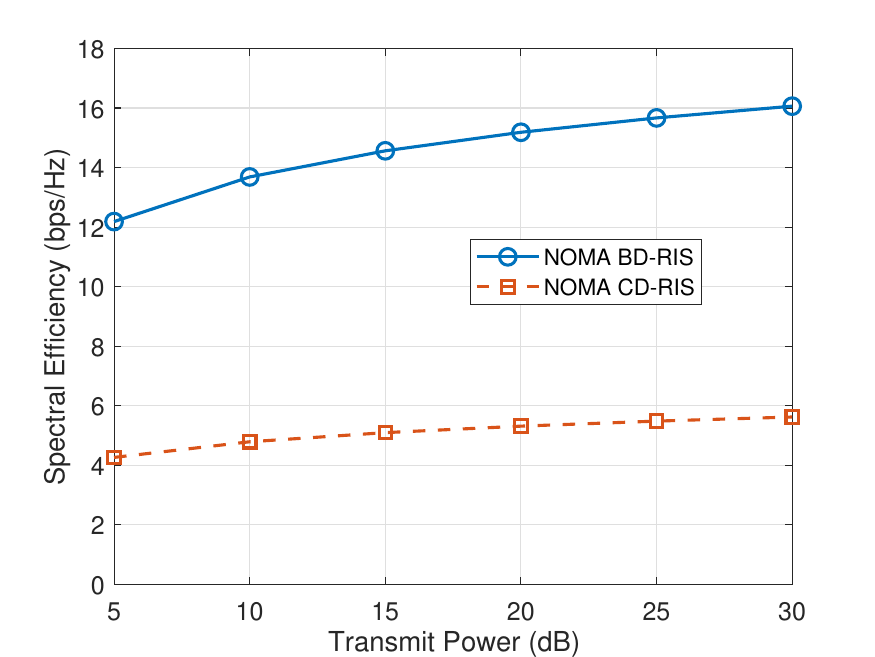}  
  \caption{}
  \label{fig:sub2}
\end{subfigure}%
\begin{subfigure}{.33\textwidth}
  \centering
  \includegraphics[height=0.2\textheight]{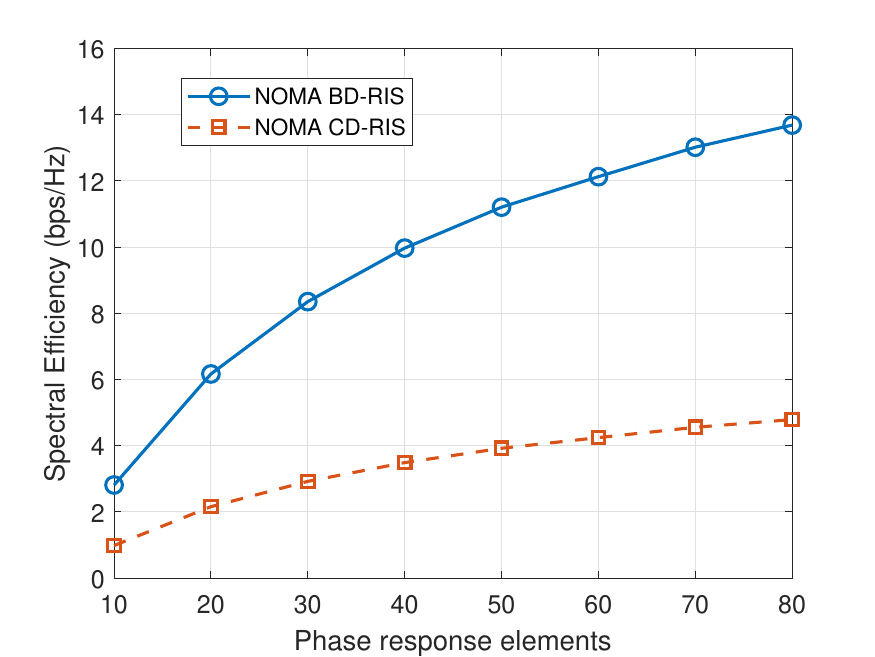}  
  \caption{}
  \label{fig:sub3}
\end{subfigure}
\caption{(a) System model of NOMA BD-RIS aided LEO communications, (b) Achievable spectral efficiency of the system versus varying transmit power of LEO, (c) Achievable spectral efficiency of the system versus varying phase response elements.}
\label{fig:test}
\end{figure*}

\subsection{System Model and Proposed Optimization}
Let us consider a downlink scenario of BD-RIS aided satellite communication, as shown in Fig. \ref{fig:test} (a), where an LEO communicates with ground users (GUs) using the power domain NOMA. In practice, the GUs operate with low directional antennas, so closing links between LEO and GUs is challenging when the network operates on a high frequency. Moreover, the direct links between LEO and GUs might not be possible, considering the mobile scenario. To compensate for path losses due to poor link budget and improve the received signal strength, we consider fully-connected reflective BD-RIS installed on the optimal location and assisting the signal delivery from LEO to GUs. Thus, GUs can receive signals from direct and BD-RIS aided links. Assuming perfect acquisition of channel state information (CSI) by channel estimation techniques, this scenario provides an upper-bound solution. In addition, the GUs employ successive interference cancellation (SIC) to decode their information from the superimposed signal. In particular, the GU with a stronger channel condition can apply SIC to remove the other GU's information before decoding its desired signal. Then, the GU with poor channel conditions cannot apply SIC and decode the signal directly by considering the other GU signal as noise.

This case study seeks to improve the achievable sum-rate of BD-RIS aided NOMA LEO satellite communication by optimizing the transmission power of LEO and PR of BD-RIS while ensuring the minimum data rate requirements of GUs. This is achieved by formulating a mathematical optimization problem for achievable sum-rate maximization which is subjected to several practical constraints such as minimum rate constraint, LEO power budget constraint, NOMA power allocation constraint, and BD-RIS PR constraint, respectively. The formulated optimization is nonlinear and NP-hard in nature due to coupling decision variables and NOMA interference. In practice, solving such problems is challenging due to high computational complexity. To make the optimization tractable, we first adopt the block descent coordinate method to split the joint optimization into two problems: 1) transmission power at LEO for a given PR at BD-RIS and 2) PR at BD-RIS, given the power allocation for LEO. Then, each problem is further transformed into linear form and simultaneously solved using a standard convex optimization such as CVX toolbox. 

\subsection{Numerical Results and Discussion} 
This section presents the numerical results to reveal the effectiveness of the proposed LEO communication scheme in the efficient allocation of NOMA transmission power and PR of the BD-RIS. The analysis is carried out by setting the transmit power to 20 dBm, the operating frequency to 3.5 GHz, and the elevation angles to $45^\circ$. The antenna gains are set to 10 dBi, with a path-loss exponent of 2.5. The dimension of the BD-RIS PREs is $0.5\,\text{m} \times 0.5\,\text{m}$, and the distance from BD-RIS to the satellite is 500 km. The altitude of the satellite is set to 600 km, and the radius of Earth is considered to be 6371 km. The reflection coefficient magnitude is 0.9, the noise power is -90 dBm,  and the Rician factor is set as 10. The simulations were conducted based on Monte Carlo simulations, where the average results were obtained over independent channel realizations. The simulation results were plotted considering the sum rate
across different numbers of BD-RIS PREs. Furthermore, the results were compared with those CD-RIS aided NOMA LEO satellite communications.
\par
 First, we investigate the impact of LEO transmit power on the spectral efficiency of the system with the number of PREs is 80, as depicted in Fig. Fig.\ref{fig:test} (b). It can be seen that the spectral efficiency of both NOMA BD-RIS and NOMA CD-RIS systems increases as the transmit power of LEO increases. We can also observe that NOMA BD-RIS aided LEO communications achieve significantly high spectral efficiency compared to its counterpart NOMA CD-RIS LEO communications. For instance, using the same system parameters when LEO is operating at 15 dB transmit power, the proposed NOMA BD-RIS provides 14.56 bps/Hz while NOMA-CD-RIS can only achieve 5 bps/Hz. This is because NOMA BD-RIS offers enhanced signal control by adopting a full scattering matrix, resulting in improved signal gain.

In addition, it is interesting to see the role of PREs on the system performance in both BD-RIS and CD-RIS systems, where the transmit power of LEO is set as 10 dB. The result in Fig.\ref{fig:test} (c) shows that as the number of BD-RIS PREs increases, the spectral efficiency of both systems also increases. This is because more PREs can effectively direct the signal towards the target users, increasing channel gain. However, the proposed NOMA BD-RIS LEO communications outperforms the NOMA CD-RIS LEO communications. The main reason for this is the advanced control of the BD-RIS system over signal manipulation compared to the CD-RIS system.

\section{Challenges and Research Directions}
\subsection{Challenges}
Although the information provided in the previous sections well establishes that BD-RIS offers significant benefits over CD-RIS, it comes with limitations that open up crucial problems needing resolution to fully exploit its advantages. On one hand, there is very limited literature highlighting the key issues associated with the use of BD-RIS in actual wireless networks and providing new research directions. On the other hand, since the use of NTN for terrestrial users is still in its infancy, the issues related to the use of BD-RIS in such settings require extensive analysis. Therefore, in this paper, we have highlighted three key issues associated with the use of BD-RIS in an NTN-assisted wireless network.

\subsubsection{\textbf{Adaptive Channel Realization}}

The accurate realization of channels is one of the key aspects that significantly supports BD-RIS in achieving its promised performance gains. The conventional channel estimation techniques used in  CD-RIS, such as semi-passive methods (enabling CD-RIS to perform pilot transmission using low-power RF chains), can easily be replicated in BD-RIS at the expense of higher power consumption. However, purely passive techniques cannot be directly adopted due to variations in RF circuitry. When incorporating BD-RIS in a NTN wireless communication scenario, the channel estimation process becomes much more complicated as the signal propagation path significantly lengthens, and fast time-varying channels are introduced (considering the LEO satellite use case of NTN). Therefore, the channel estimation techniques used in CD-RIS cannot be directly adapted to the BD-RIS aided NTN scenario. Instead, innovative upgrades and/or new techniques are required that not only improve the accuracy of channel estimation but also support low power consumption, promoting sustainability.

\subsubsection{\textbf{RF Circuitry Constraints}}

The development of BD-RIS is still in its early stages, where the mathematical models used to simulate it for various applications mostly rely on ideal assumptions, such as lossless impedance components, perfect matching, and no mutual coupling between BD-RIS PREs. These assumptions directly impact the BF capability of BD-RIS by increasing the complexity of performing BF. Moreover, they also limit the implementation of the BD-RIS matrix using discrete values, while relying on simple quantization mechanisms. Therefore, it is necessary to consider these impairments when designing BD-RIS. For BD-RIS aided NTN applications, these RF circuitry constraints present additional challenges. Signals from NTN systems travel significant distances and face atmospheric turbulence, so relying on ideal models in two-way communication significantly impacts BD-RIS aided NTN communication. To fully exploit the benefits of BD-RIS in NTN-assisted scenarios, RF circuitry constraints need to be fully incorporated into mathematical models and simulation analyses using innovative methods.

\subsubsection{\textbf{Receiver Sensitivity}}

One of the main concerns in NTN-assisted terrestrial users is the distance traveled by the wireless signal between the transmitter and the receiver. Although there is supposed to be an acceptable quality of LoS connection between a satellite and a ground user, the free space path loss and other atmospheric turbulence significantly degrade the signal quality. To fully exploit the benefits of BD-RIS in an NTN scenario, the sensitivity of BD-RIS needs to be effective enough to differentiate between noise and the NTN signal. In CD-RIS, significant attention has been given to overcoming this issue. For instance, in \cite{liang2021angle}, the structure of CD-RIS is repurposed to address angular sensitivity by utilizing metallic vias to develop programmable meta-atoms that are insensitive to angular distortions. A similar setup can be adopted for BD-RIS, but when considering NTN-assisted BD-RIS wireless communication scenarios, this issue needs to be addressed with innovative designs and algorithms.

\subsection{Research Directions}

\subsubsection{\textbf{Artificial Intelligence/Machine Learning}} 

On one hand, as elucidated in the preceding subsections, the three predominant challenges—namely adaptive channel realization, constraints inherent to RF circuitry, and receiver sensitivity—impose significant limitations on the comprehensive adaptation of BD-RIS within the context of NTN. On the other hand, forthcoming inquiries into BD-RIS-enhanced NTN ought to concentrate on mitigating these challenges through the integration of cutting-edge methodologies. Artificial Intelligence and Machine Learning can serve a pivotal function in the optimization of BD-RIS by facilitating real-time adjustments of signal reflections, adjusting RF circuitry losses, forecasting channel fluctuations, and implementing self-optimization based on feedback mechanisms, thereby augmenting performance in dynamic operational environments.
\subsubsection{\textbf{Physical Layer Security}}
The advancement of physical layer security algorithms is paramount for the protection of sensitive communications within NTN, wherein the inherently broadcast nature of satellite-based systems renders them susceptible to unauthorized access and signal manipulation. By fortifying these algorithms, BD-RIS can enhance the security of communications, refine real-time signal reflection adjustments, and adapt to changing network conditions, thereby ensuring more reliable and efficient performance in prospective NTN scenarios.
\subsubsection{\textbf{Joint Communications and Sensing}}
Joint communication and sensing are promising research frontiers in NTNs due to the dual benefit of using a single platform for monitoring and data transmission. However, accurate signal management and the possibility of communication and sensor signal interference make this integration difficult, especially in dynamic and resource-constrained environments like space. BD-RIS improves signal directionality and intensity, enabling precise beamforming that adapts to changing conditions, which protects communication and sensor signals. The joint communications and sensing applications in NTNs with BD-RIS hold significant potential and warrant further exploration.

\section{Conclusion}
This article proposed BD-RIS for 6G NTNs to further enhance control over the radio environment with a beyond diagonal PRMs. First, we provided the fundamentals of 6G NTNs and discussed the atmospheric layer and space layer, followed by various hardware architectures, modes and advantages of BD-RIS technology. Then, we discussed different applications of BD-RIS in 6G NTNs. To investigate the performance of BD-RIS in NTNs, we also provided a case study on BD-RIS enabled NOMA LEO satellite communication and compared its performance with NOMA CD-RIS LEO communications. Finally, we studied challenges and future research directions in BD-RIS enabled NTNs and concluded our work.

\ifCLASSOPTIONcaptionsoff
  \newpage
\fi

\bibliographystyle{IEEEtran}
\bibliography{Wali_EE}

\section*{Biographies}\small
\noindent {\bf Wali Ullah Khan [M]} (waliullah.khan@uni.lu) received a Ph.D.
degree in information and communication engineering from
Shandong University, Qingdao, China, in 2020. He is currently
working with the SIGCOM Research Group, SnT, University of Luxembourg.
\vspace{0.2cm}

\noindent {\bf Asad Mahmood [S]} (asad.mahmood@uni.lu) received
his Master degrees in Electrical Engineering from COMSATS University Islamabad, Wah Campus, Pakistan. He is currently pursuing the Ph.D.
degree with the Interdisciplinary Centre for Security,
Reliability, and Trust (SnT), University of Luxembourg.
\vspace{0.2cm}

\noindent {\bf Muhammad Ali Jamshed} (muhammadali.jamshed@glasgow.ac.uk) received a Ph.D. degree from the University of Surrey, Guildford, U.K, in 2021. He is currently working with the James Watt School of Engineering, University of Glasgow, UK. 
\vspace{0.2cm}

\noindent {\bf Eva Lagunas [SM]} (eva.lagunas@uni.lu) received a Ph.D. degree
in telecommunications engineering from the Polytechnic University of Catalonia (UPC), Barcelona, Spain, in 2014. She currently holds a research scientist position in the SIGCOM Research
Group, SnT, University of Luxembourg.
\vspace{0.2cm}

\noindent {\bf Manzoor Ahmed} (manzoor.achakzai@gmail.com) received his Ph.D.
degree from Beijing University of Posts and Telecommunications, Beijing China, in 2015. He is currently a professor in the School of Computer and Information Science and also with the Institute for AI Industrial Technology Research, Hubei Engineering University, Xiaogan City, 432000, China
\vspace{0.2cm}

\noindent {\bf Symeon Chatzinotas [F]} (symeon.chatzinotas@uni.lu) received
Ph.D. degrees in electronic engineering from the University of Surrey, Guildford, United Kingdom, in 2009. He is currently a full professor or Chief Scientist I and the co-head of the SIGCOM Research Group, SnT, University of Luxembourg.

\end{document}